\documentclass[bibyear]{aa}  
\usepackage{txfonts}
\usepackage{url}
\usepackage{enumitem}
\usepackage{graphicx}

\def\spose#1{\hbox to 0pt{#1\hss}}
\newcommand\lsim{\mathrel{\spose{\lower 3pt\hbox{$\mathchar"218$}}
     \raise 2.0pt\hbox{$\mathchar"13C$}}}
\newcommand\gsim{\mathrel{\spose{\lower 3pt\hbox{$\mathchar"218$}}
     \raise 2.0pt\hbox{$\mathchar"13E$}}}

\def\ltsima{$\; \buildrel < \over \sim \;$}
\def\lsim{\lower.5ex\hbox{\ltsima}}
\def\gtsima{$\; \buildrel > \over \sim \;$}
\def\gsim{\lower.5ex\hbox{\gtsima}}

\def\apjl{ApJL}
\def\apjs{ApJSS}

\def\aap{A\&A}

\usepackage[normalem]{ulem}

\begin{document}

\title{Cross section of curvature radiation absorption }
\titlerunning{Cross section of curvature radiation absorption}
\authorrunning{Locatelli \& Ghisellini}
\author{Nicola Locatelli\inst{1,2,3,4}\thanks{E--mail: locatelli@ira.inaf.it},
Gabriele Ghisellini\inst{1} \\
}
\institute{
$^1$ INAF -- Osservatorio Astronomico di Brera, via E. Bianchi 46, I-23807 Merate, Italy.\\
$^2$ Univ. di Milano Bicocca, Dip. di Fisica G. Occhialini, Piazza della Scienza 3, I--20126 Milano, Italy.\\
$^3$ Univ. Alma Mater Studiorum, Dip. di Fisica e Astronomia DIFA, via Piero Gobetti 93/2, 40129-Bologna, Italy.\\
$^4$ Istituto di Radio Astronomia (IRA) sede di Bologna, via Piero Gobetti 101, 40129 Bologna, Italy.
}

\abstract{
When treating the absorption of light, it is instructive to focus on the absorption 
coefficient related to the probability of photons to survive while traversing a layer
of material. 
From the point of view of particles doing the absorption, however, the elementary
interaction of the particle with the photon is best described by the corresponding
cross section.
We revisit curvature radiation in order to find the absorption cross 
section for this process, making use of the Einstein coefficients and their relations 
with spontaneous and stimulated emission and true absorption.
We derive the cross section as a function of the emission angle $\psi$ (i.e. the angle between
the instantaneous velocity vector and the direction of the photon) and the 
cross section integrated over angles. 
Both are positive, contrary to the synchrotron case  for which the cross section
can be negative for large $\psi$.
Therefore, it is impossible to have curvature radiation masers. 
This has important consequences for sources of very large brightness temperatures
that require a coherent emission process, such as pulsars and fast radio bursts.
}
\maketitle

\section{Introduction}

When dealing with the absorption process in general, we are usually interested
in the absorption coefficient, namely the resulting intensity 
and spectrum of the radiation.
But there is a complementary view, namely the electron point of view. In this case we are interested in the absorbing electron
gain or loss of energy and momentum as a result of the total absorption rate, which includes
stimulated emission.
The latter view is best captured by the concept of cross section.
As in the synchrotron case, this helps to treat the problem of energy
exchange between particles (both protons and electrons) eventually leading to their
thermalization.
If the absorbed radiation carries some linear momentum, a fraction of this momentum
lets the absorbing particles acquire a pitch angle and emit
by synchrotron radiation, while the component along the magnetic field
line accelerates the absorbing particle in that direction.

Curvature radiation is very similar to synchrotron radiation, in which
the Larmor radius takes the place of the curvature radius of the magnetic field
(see e.g. Jackson 1962). 
The absorption cross section of synchrotron radiation, in some special cases,
can become negative, and therefore a synchrotron maser
is possible (Ghisellini \& Svensson 1991,  hereafter GS91).
Given the similarities between the synchrotron and curvature emission
processes, we wonder if a curvature maser is possible.
This would have a great impact on the studies of the emission process
of fast radio bursts (FRBs), helping to explain the large observed
brightness temperatures, requiring coherent radiation.

Jackson (1962, see also the 1999 edition) gives the
classical treatment for curvature emission, but not absorption.
Coherent absorption by a bunch of particles was considered by 
Cocke and Pacholczyk (1975).
Blandford (1975) found that the absorption coefficient is positive in general,
but can be negative in specific cases and geometries.
However, Melrose (1978) argued against this conclusion, finding the impossibility
of a curvature maser even in the presence of ambient particles.
Later, Zheleznyakov \& Shaposhnikov (1979) found instead results in agreement to Blandford (1975).
%This latter study focuses however on a specialized scenario (neutron stars with magnetic field
%that can be greater than the critical value $B_{\rm c}\sim 4.4 \times 10^{13}$ G,
%when Landau levels become important).
The issue is therefore not completely clear and we would like to explain the 
absorption process at the elementary level, resorting to the Einstein coefficients
and their relations.
In other words, we would like to compute the probability for 
true absorption or stimulated emission of the single electron
of a given energy when interacting with a photon of energy $h\nu$.

We leave the construction of a more complex scenario, 
involving a realistic dipole field around a neutron star, its rotation,  and
the acceleration of particles along divergent field lines, to a future paper
concerning pulsars (see the review by e.g. Usov 2000 and Lyutikov, Machabeli \& Blandford 1999;
Lyutikov, Blandford \& Machabeli 1999).
Although our main aim is to find the cross section for curvature absorption, we
revisit concepts and formulae that are well known, but that we re-use with 
our notation.

\section{Curvature cross section}

When deriving the absorption cross section, we are dealing with a
particle at some energy level  $\gamma mc^2$  (for simplicity, we call this level 2) 
and an incoming photon of energy $h\nu$.
The particle can absorb it, jumping to a higher energy level (level 3, with energy
$\gamma mc^2+h\nu$), or can emit another photon of same energy, phase, and direction
of the incoming photon, through stimulated emission.
In this case it jumps to level 1 (of energy $\gamma mc^2-h\nu$).
Both these processes are related to the spontaneous process of emitting
a photon of the same energy $h\nu$, through the Einstein coefficients
\begin{equation}
B_{21}=\frac{c^2}{2h\nu^3} A_{21}; \quad B_{23} = B_{32} =    \frac{c^2}{2h\nu^3} A_{32}
,\end{equation}
where $B_{21}$ and $B_{32}$ are the Einstein coefficients for stimulated emission,
while $B_{12}$ and $B_{23}$ are the coefficients for true absorption, and $A_{21}$ and
$A_{32}$ correspond to spontaneous emission.
The probability that the particle initially in level 2 absorbs the photon
($B_{23}$) depends on the spontaneous emissivity in level 3 ($A_{32}$),
while the probability to have induced emission ($B_{21}$) depends on the spontaneous
emissivity in level 2 ($A_{21}$).

We set $\epsilon\equiv h\nu/(m c^2)$, and measure the particle energy
in units of $m c^2$, and momentum $p$ in units of $m c$. 
We then have $\gamma_1 = \gamma_2-\epsilon$ and $\gamma_3=\gamma_2+\epsilon$.

Consider the emissivity of the single particle (in erg s$^{-1}$ Hz$^{-1}$ ster$^{-1}$) 
for a small emission angle $\psi$  between the (instantaneous) velocity of the particle
and the direction of the produced photon.
At the two energies $\gamma_2$ and $\gamma_3=\gamma_2+\epsilon$ the emissivity can be written as
%:
\begin{eqnarray}
j(\nu, \gamma_2, \psi) \,  &=&  \, h\nu A_{21} \nonumber \\
j(\nu, \gamma_2+\epsilon, \psi) \, &=& \, h\nu A_{32}
.\end{eqnarray} \label{js}
%
%Where $4\pi \gamma_1 p_1$ and  $4\pi \gamma_2 p_2$ are the terms associated to the phase space. 
%Note that the emissivity depends upon the phase space of ``arrival"
%(namely the one corresponding to the energy of the electron after the transition).
%
The differential cross section for true absorption ($d\sigma_{\rm ta}/d\Omega$) 
and stimulated emission  ($d\sigma_{\rm se}/d\Omega$) can then be
written as (see GS91)
\begin{equation}
\frac{d\sigma_{\rm ta}}{d\Omega} \, =\, h \nu B_{23} 
\,=\,  \frac{c^2 }{2h\nu^3} j(\nu, \gamma_2+\epsilon, \psi)
\label{ta}
\end{equation}
\begin{equation}
\frac{d\sigma_{\rm se}}{d\Omega} \, =\, h \nu B_{21}
\,=\,  \frac{c^2}{2h\nu^3}  j(\nu, \gamma_2, \psi)
\label{se}
.\end{equation}
We note that the above formulae implicitly assume that the momentum of the particle, after
having absorbed the photon, is still along the magnetic field line, even for
$\psi \ne 0$.
% assume that, at first order, all the energy of the photon is converted in 
% particle's momentum along the direction of motion. 
This is different from the synchrotron case, 
for which a phase space factor ($\propto \gamma p$) given by all the possible final states 
of the momentum of the free particle should also be considered in equations~\ref{js}-\ref{stot}, 
(see GS91 for comparison).
The net absorption cross section is the difference between the two, very similar,
cross sections. 
When $\epsilon\ll\gamma$  we can write
\begin{equation}
\frac{d\sigma }{ d\Omega} \, =\, \frac{d\sigma_{\rm ta} }{d\Omega} - \frac{d\sigma_{\rm se} }{ d\Omega} = \frac{1}{2m\nu^2} \frac{\partial}{\partial\gamma}\left[ j(\nu, \gamma, \psi) \right]
\label{stot}
.\end{equation}
The differential cross sections in Eq. \ref{ta} and Eq. \ref{se} refer to one particular
direction $\psi$ of the incoming photon.
This direction coincides with the corresponding emission angle for the emissivity.
If the particles emits a photon in front of it, the corresponding cross section
is for an incoming photon at the same angle (i.e. the photon is coming from
behind the particle).
In general, if the energy of the particle increases, so does its emissivity.
%This is why, in general, the total (i.e. integrated over all angles) cross section is positive.

In the synchrotron case, however, there are special cases where the emissivity for
specific directions decreases when the particle energy is increased.
This occurs when the incoming photon arrives at an angle larger than
the characteristic beaming angle $1/\gamma$.
GS91 have shown that when $\psi > 1/\gamma,$ the increase of $\gamma$ makes the synchrotron emissivity decrease, possibly even more than the increase of the phase space factor 
$\gamma p$, which multiplies the emissivity in the synchrotron case. 

The stimulated emission becomes larger than the true absorption, the total
cross section becomes negative, and there is the possibility to have a maser or laser.

% ============================================================================================

\begin{figure} 
\vskip -2.9 cm
\hskip -0.7 cm
\includegraphics[width=10cm]{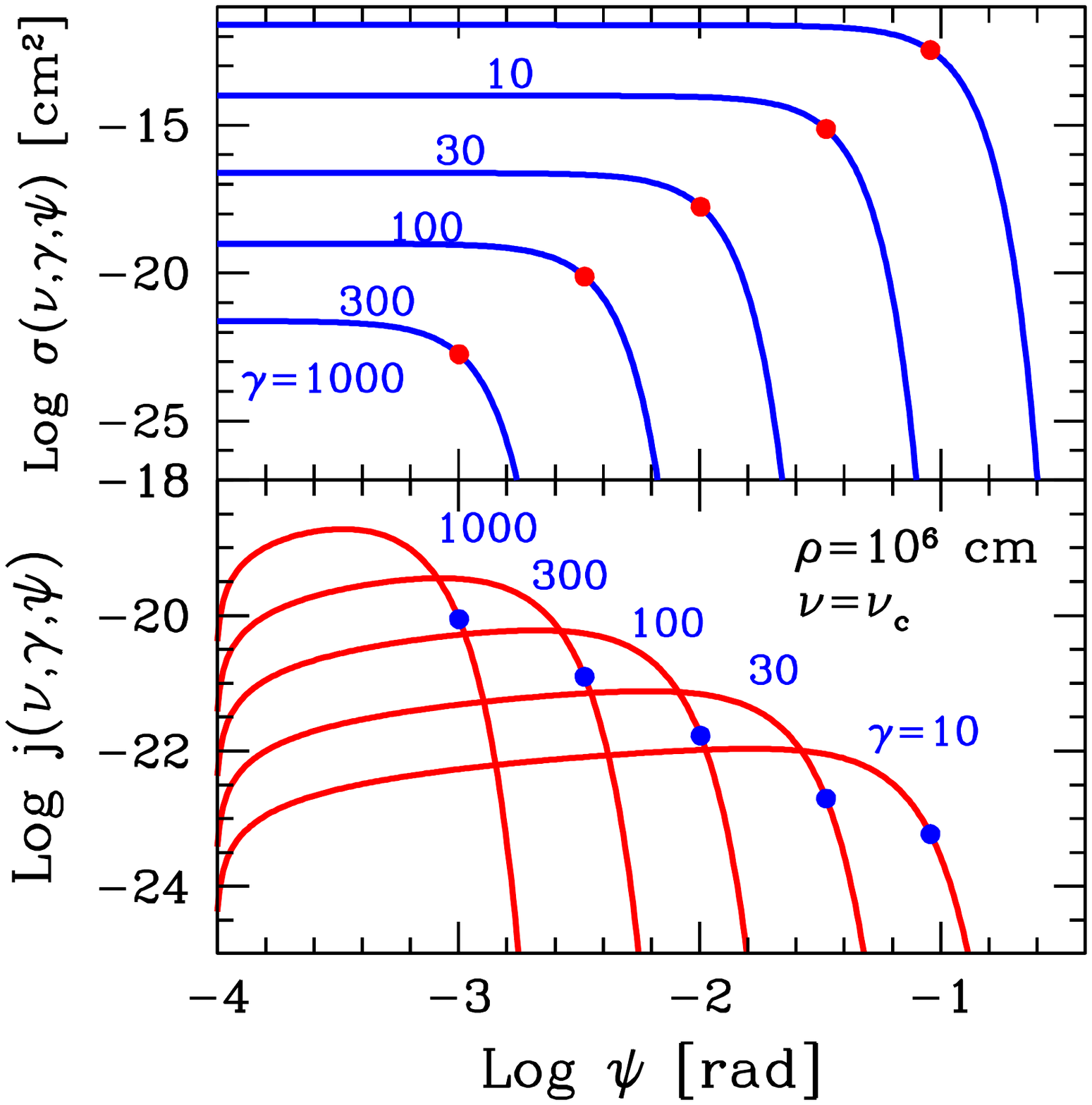} % ,width=12cm}
\vskip -2.8 cm
\caption{Top panel: the curvature absorption cross section as a function of the emission
angle $\psi$ for different values of $\gamma$. 
We assume a curvature radius $\rho=10^6$ cm and a frequency $\nu=\nu_{\rm c}=(3/2)\nu_0\gamma^3$.
Bottom panel: single electron emissivity as a function of $\psi$
for different values of $\gamma$, as labelled.
The circles in both panels correspond to $\psi=1/\gamma$. 
} 
\label{sigmapsi}
\end{figure}

Following Jackson (1999), we report the single particle emissivity for curvature radiation
as a function of the emission angle $\psi$.
First let us introduce the notation%
\begin{eqnarray}
\rho\, &=&\, {\rm curvature \, radius}\nonumber \\
\nu_0 &\equiv& {c\over 2\pi \rho   }; \quad \nu_{\rm c} \equiv  \frac{3}{2}  \gamma^3 \nu_0; \nonumber \\
x&\equiv&  \frac{\nu}{\nu_{\rm c}};\qquad t \equiv \psi^2\gamma^2; \nonumber \\
y &\equiv& \frac{ 2\nu}{3 \gamma^3 \nu_0 } (1+t)^{3/2} = \frac{\nu}{\nu_{\rm c}}(1+t)^{3/2} 
= x(1+t)^{3/2}
\label{def}
.\end{eqnarray}
Then we have
\begin{eqnarray}
j_{\rm c}(\nu, \gamma,\psi)  &=& \frac{3}{4\pi^2} \, \frac{ e^2}{\rho } x^2 \gamma^2\, (1+t) \nonumber \\
   &\times& \left[ (1+t) K^2_{2/3}(y) +t K^2_{1/3}(y)  \right]
\label{emisspsi}
,\end{eqnarray}
which is valid for $\gamma\gg 1$ and $\psi\ll 1$. 
$K_a(y)$ is the modified Bessel function of order $a$.
It is instructive to compare this emissivity with the synchrotron emissivity, i.e.
\begin{eqnarray}
j_{\rm s}(\nu, \gamma,\psi)  &=& {3 \over  4\pi^2} \, { e^2 \over r_{\rm L} } x_{\rm s}^2 \gamma^2\, (1+t) \nonumber \\
   &\times& \left[ (1+t) K^2_{2/3}(y_{\rm s}) +t K^2_{1/3}(y_{\rm s})  \right]
,\end{eqnarray}
where $r_{\rm L}$ is the Larmor radius, $r_{\rm L} = \gamma \beta m c^2/( eB)$,
$x_{\rm s}=\nu/\nu_{\rm s}= 2\nu/(3\gamma^2 \nu_{\rm L}\sin\theta)$, and
the Larmor frequency $\nu_{\rm L}=eB /(2\pi mc)$.
The argument of the Bessel functions $y_{\rm s}$ is the same as in 
Eq. \ref{def}, but with $x_{\rm s}$ replacing $x$. 
As expected, the functional forms are the same.
We note that
\begin{enumerate}
\item
the curvature emissivity does not depend  on the mass of the particle, 
contrary to the synchrotron emissivity (via the Larmor radius and frequency);

\item the typical frequency for curvature radiation is 
$\nu_{\rm c}=(3/2)\gamma^3 c/(2\pi \rho)\propto \gamma^3$.
Here $\rho$ is independent of $\gamma$, contrary to the 
Larmor radius that is instead 
$\propto \gamma$, which makes $\nu_{\rm s}=(3/2)\gamma^3 c/(2\pi r_L)\propto \gamma^2$. 
The overall dependence from the particle energy is thus different from the
curvature to the synchrotron case.

\end{enumerate}
Following Eq. \ref{stot} the differential cross section is

\begin{eqnarray}
{d\sigma_{\rm c} \over d\Omega} (\nu, \gamma, \psi)  
&=&  {4 \over 3 } {e^2\over m c^2}  \, {\rho\over \gamma^5 } \,\,
 \{ 3y [(1+t)K_{2/3}(y)K_{5/3}(y)       \nonumber \\
&+ & t K_{1/3}(y)K_{4/3}(y) ] -4(1+t) K^2_{2/3}(y) \nonumber \\
&- &   2tK^2_{1/3}(y) \}  
 \label{ds}
 \end{eqnarray}
As shown below, this cross section is  positive for all values of $\psi$.
Fig. \ref{sigmapsi} shows ${d\sigma_{\rm c} /d\Omega}$ 
as a function of $\psi$ for different values of $\gamma$.
For each curve we assume $\nu=\nu_{\rm c}=(3/2)\nu_0\gamma^3$.
The circles correspond to $\psi=1/\gamma$, i.e. $t=1$. 

%--------------------------------------------------
%--------------------------------------------------

%--------------------------------------------------
\begin{figure} 
\vskip -0.3 cm
\hskip -0.3 cm
\includegraphics[width=9.5cm]{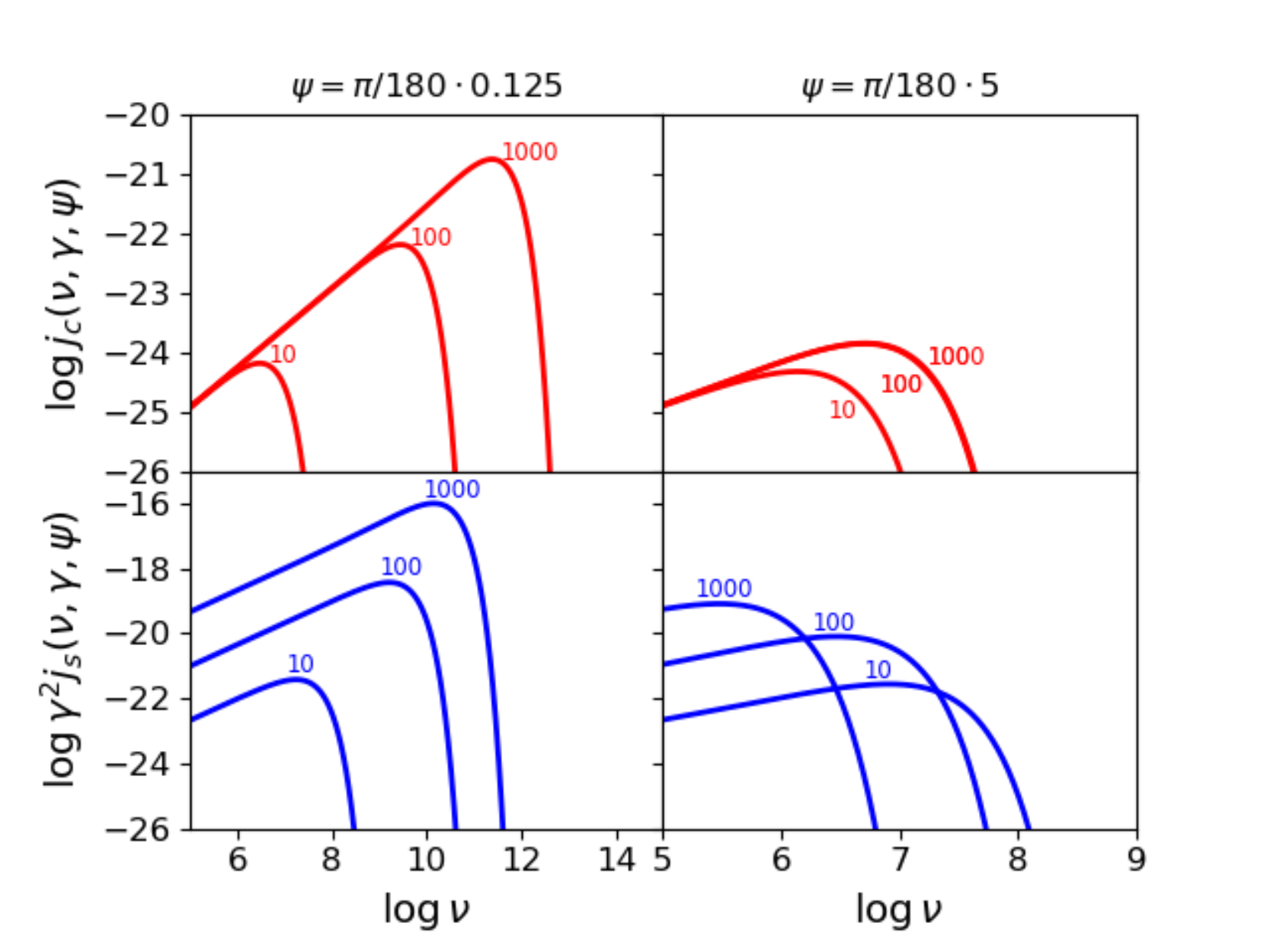} % ,width=12cm}
% \vskip -0.3 cm
\caption{Top panels: the curvature emissivity 
as a function of frequency $\nu$ for different $\gamma$ (labelled) 
and for two values of the emission angle $\psi$,
$0.125^\circ$ (left column) and $5^\circ$ (right column). 
We can see that at large $\psi$ the emission at high frequencies is quenched, but 
that the emissivity at larger $\gamma$ is always greater than the emissivity at lower $\gamma$.
Bottom panels: the same, multiplied by the phase space factor $\gamma p$, 
for the synchrotron case. 
Contrary to the curvature case, for large $\gamma$ the emissivity {\it can be smaller}
than the one at lower $\gamma$. A magnetic field $B=1$ G 
was considered in calculations for comparison.
% Bottom panels: the peak frequency of emission as a function of $\gamma$ for curvature (blue line)
% and synchrotron (red) emission.
% While the peak emission of curvature radiation remains constant at large $\gamma$,
% in the synchrotron case it decreases.
} 
\label{nomaser}
\end{figure}
%--------------------------------------------------
%--------------------------------------------------
\begin{figure} 
\vskip -2.9 cm
\hskip -0.7 cm
\includegraphics[width=10cm]{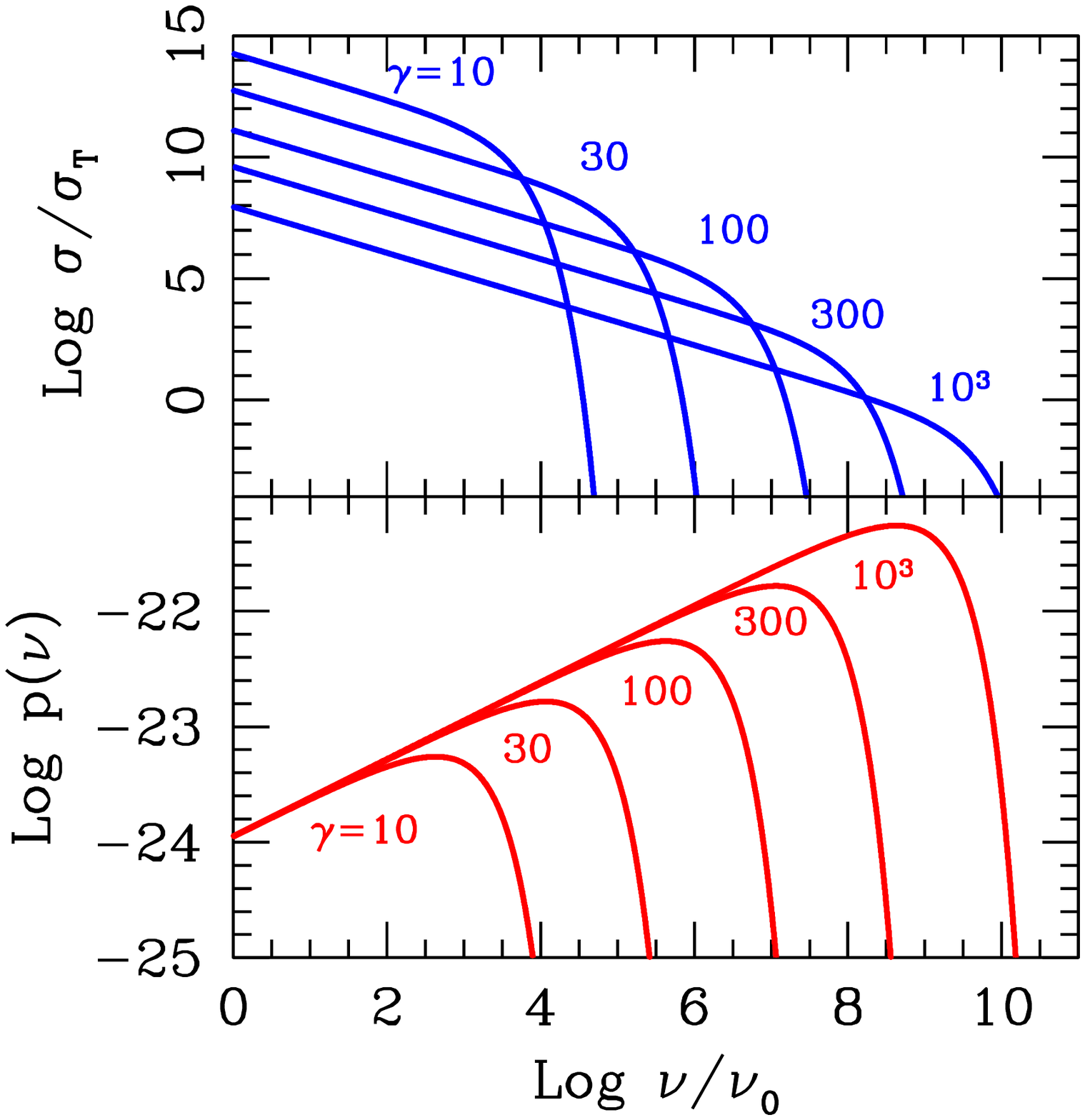} % ,width=12cm}
\vskip -2.8 cm
\caption{Top panel: the curvature absorption cross section, normalized to the scattering Thomson cross
section $\sigma_{\rm T}$, as a function of $\nu/\nu_0$ for different values of $\gamma$.
A curvature radius $\rho=10^6$ cm has been assumed.
For $\nu/\nu_0=1$, $\sigma_{\rm c}$ is of the order of $\rho/(r_{\rm e}\gamma)$ (namely
a factor $\sim 10^{19}/\gamma$) larger than $\sigma_{\rm T}$.
Bottom panel: single electron emissivity, again as a function of $\nu/\nu_0$ 
for different values of $\gamma$, as labelled.
} 
\label{sigma}
\end{figure}
%--------------------------------------------------
% -------------------------------------------
%\begin{figure} 
%%\vskip -2.9 cm
%%\hskip -0.7 cm
%\includegraphics[width=8cm]{emission_cone.png} % ,width=12cm}
%%\vskip -2.8 cm
%\caption{Cartoon illustrating the physical interpretation of the cross section.}
%\label{geometry}
%\end{figure}
% -------------------------------------------

\subsection{Impossibility of curvature maser}

To understand why ${d\sigma_{\rm c}/d\Omega}$ is always positive, we
consider the emissivity (Eq. \ref{emisspsi})
as a function of $\nu$, as shown in the top and middle panels of Fig. \ref{nomaser}.
The emission angles are $0.125^\circ$ (left) and $5^\circ$ (right).
The plotted curves correspond to different values of $\gamma$, as labelled.
The top panels shows that increasing $\gamma$, the emissivity increases 
both for small and large angles $\psi$. 
This contrasts the behaviour of the synchrotron emissivity,
shown in the bottom panels. In this case, for large angles $\psi>1/\gamma$,
and at high frequencies, the emissivity {\it decreases} increasing $\gamma$.
%This is true even if we multiply the emissivity by the phase space factor $\gamma^2$.
For small $\psi$, the behaviour is similar to the curvature radiation.
% The bottom panels compare how the peak frequency of the emission changes as a function of $\gamma$ 
% for curvature (blue dashed line) and synchrotron (red solid line) emission.
% At large $\gamma$ (i.e. outside the beaming cone $1/\gamma$) 
% the peak frequency $\nu_{\rm peak}$ of curvature radiation remains constant, while
% it decreases in the synchrotron case. 
% This implies that, increasing $\gamma$, the observer sees the same 

The reason for this difference lies in the difference between the Larmor radius,
proportional to $\gamma$, and the curvature radius, that is constant.
This is the reason why the total power emitted by the single particle (integrated over angle)
is $\propto \gamma^2$ for synchrotron and $\propto \gamma^4$ for curvature emission.
The latter dependence is so strong that at large $\psi$ the received radiation is
still larger for increasing $\gamma$, even if the beaming cone (of semi-aperture angle $1/\gamma$)
becomes narrower.

\subsection{Total cross section}

The emissivity  integrated over the emission angles (see e.g. Jackson 1999)
gives the power per unit frequency
\begin{equation}
p_{\rm c}(\nu,\gamma)\, =\, 2\pi \sqrt{3} {e^2 \over c} \nu_0 \gamma x \int_x^\infty K_{5/3}(x^\prime) dx^\prime
\label{psingle}
.\end{equation}
We can then find the cross section integrated over angles using, in Eq. \ref{stot},
the above emissivity instead of $j(\nu,\gamma,\psi)$.
We obtain
\begin{equation}
\sigma_{\rm c} (\nu,\gamma)  \,  = \,  
     {1  \over 2 \sqrt{3}}\, {e^2\over mc^2} \,  {\rho\over \gamma^6 } \left[ K_{5/3}(x) - \frac{2}{3x} \int_x^\infty K_{5/3}(y)\,dy \right]    
% \nonumber  \\
% {\sigma_{\rm c} (\nu,\gamma) \over \sigma_{\rm T}} \, &=&\,  
%      {\sqrt{3}  \over 12} {m_{\rm e}\over m}  \, {\rho\over r_{\rm e}}{1\over \gamma^6 } K_{5/3}(2x)  
 \label{sigmatot}
 .\end{equation}
The top panel of Fig. \ref{sigma} shows the behaviour of $\sigma_{\rm c} (\nu,\gamma)$  
as a function of $\nu/\nu_0$ for different values of $\gamma$, while the bottom panel 
shows the corresponding power emitted by the single particle.

\subsubsection{Geometrical interpretation}

We can find the asymptotic behaviour of  of Eq. \ref{sigmatot} for low frequencies (i.e. $x \ll 1$),
using $K_a(x) \to 2^{a-1} \Gamma (a) x^{-a}$.
Furthermore, let us consider electrons, and thus $e^2 /(m_{\rm e}c^2) $ is equal to the classical 
electron radius $r_{\rm e}$.
In this case, 
\begin{equation}
\sigma_{\rm c} (\nu,\gamma)  \,  = \,  
     \frac{\sqrt{3}}{2^{4/3}}\,  \Gamma\left(\frac{5}{3}\right)  \, {r_{\rm e} \rho\over \gamma} 
     \left({\nu\over \nu_0}\right)^{-1} , \qquad {\nu \ll \nu_{\rm c}}
 .\end{equation}
For $\nu=\nu_0$, we have 
\begin{equation}
\sigma_{\rm c} (\nu_0,\gamma)  \, \approx \,  \, {r_{\rm e} \rho\over \gamma } 
 .\end{equation}
This is the maximum value of the cross section and it is equivalent to a physical surface 
of the order of the classical electron radius times the curvature radius divided by $\gamma$. 
We note that $\rho/\gamma$ is the actual segment of arc travelled by the electron illuminating
the observer.

\section{Self absorption frequency and brightness temperature}

In the Appendix we derive the emissivity and absorption coefficient $\alpha_{\rm c}(\nu)$
for a particle distribution $N(\gamma) = N_0 \gamma^{-n}$
between $\gamma_{\rm min}$ and $\gamma_{\rm max}$.
We can then derive the self-absorption frequency for
a power law distribution of particles emitting and absorbing
curvature radiation.

The absorption optical depth of a layer of length $R$ is
$\tau_\nu = R\alpha_{\rm c}(\nu)$.
The self-absorption frequency $\nu_{\rm t}$ is defined through $\tau_{\nu_{\rm t}}=1$. 
For $N(\gamma)=N_0\gamma^{-n}$, with $\gamma_{\rm min}\sim 1$, we have
\begin{equation}
\nu_{\rm t}  \, = \, \nu_0\left[ C(n){e^2\over mc^2} \rho R N_0 \right]^{3/(n+5)} \label{cn}
,\end{equation}
where \begin{equation}
C(n) = \frac{3^{(2n-5)/6}}{16} \,n \frac{n+4}{n+2}  \Gamma\left(\frac{n}{6}\right)  \Gamma\left(\frac{n+4}{6}\right) \nonumber.
\end{equation}
Typical values for $C(n)$ are $C(1)=0.38$, $C(2)=0.42$, $C(3)=0.52$, $C(4)=0.7$.

Using $\rho=R=10^6$ cm, $n=3,$ and $N_0=10^{15}$ cm$^{-3}$,
we derive $\nu_{\rm t}\sim 1$ GHz.
The large particle density is just a fraction of the order of the Goldreich \& Julian (1969) density $n_{\rm GJ}$
close to the surface of a rapidly spinning magnetar,%
\begin{equation}
N_{0,\rm GJ} \, \sim \, 0.07 {B\over P_{\rm ms}}\, = \, 
7\times 10^{15} \, {B_{14} \over P_{\rm ms} } \,\,\, {\rm cm^{-3}}
\label{ngj}
,\end{equation}
where $P_{\rm ms}$ is the period of the magnetar measured in milliseconds. 

The brightness temperature is defined by equating the monochromatic intensity 
to the Raleigh--Jeans part of the black-body intensity
\begin{equation}
I_\nu \, = \, 2 k_{\rm B} T_{\rm B} {\nu^2\over c^2}
.\end{equation}
In our case, $T_{\rm B}$ is maximized at the self-absorption frequency $\nu_{\rm t}$.
Using the source function (Eq. \ref{source} of the Appendix) at $\nu_{\rm t}$
and derive
\begin{eqnarray}
T_{\rm B,max} \, &=&\, { S_{\rm c}(\nu_{\rm t}) \over 2 k_{\rm B} } \,  {c^2\over \nu^2_{\rm t}}
\nonumber \\
\, &=&  {2\pi^2 D(n)\over C(n) ^{(n+4)/(n+5)}  }
 {mc^2\over k_{\rm B}} \left[ {e^2\over mc^2 } \rho R N_0 \right]^{1/(n+5)}
\label{tb}
,\end{eqnarray}
%
%\textbf{Specific values are $A(2)=0.93$, $A(3)=0.28$, $A(4)=0.12$.
where $D(n)$ is given by Eq. \ref{emiss}.
With $R=\rho=10^6$ cm and $n=3$, we have 
$T_{\rm B, max} = 1.8\times 10^{12} (N_0/10^{15}\, {\rm cm^{-3}})^{1/8}$ K.

\section{Conclusions}

We have revisited the process of curvature radiation in order to calculate the
absorption cross section between the particle and an incoming photon.
In general, the concept of cross section allows us to consider the basic process
of the particle--photon interaction from the point of view of the electron.
Therefore it can be useful when considering the momentum and
energy gained by the particles absorbing radiation through the 
radiative process under consideration. 
All this has been already considered for the synchrotron process (by e.g. GS91), 
but it was never derived before for curvature radiation. 

The derived cross section can be several orders of magnitude larger than the scattering Thomson
cross section.
Thus the absorption of relatively low energy photons by a generic electron
moving along a magnetic field line may easily become the leading photon--particle
interaction. 
Exchange of energy between different emitting and absorbing particles
proceeds via the exchange of photons, and this can thermalize the particles
even in the absence of Coulomb collisions in rarefied, hot and magnetized 
plasma, similar to the thermalization that can occur in synchrotron sources
(Ghisellini, Haardt \& Svensson 1998).
Comparing the  cross sections for Thomson scattering, synchrotron absorption,
and absorption of curvature radiation we have
\begin{eqnarray}
\sigma_{\rm T} & \propto & r^2_{\rm e},\quad \, \qquad h\nu< m_{\rm e}c^2 \nonumber \\
\sigma_{\rm S} & \propto & {r_{\rm e} r_{\rm L}\over \gamma},\qquad  \nu = {\nu_{\rm L}\over \gamma} \nonumber \\
\sigma_{\rm c} & \propto & {r_{\rm e} \rho \over \gamma}, \,\, \qquad  \nu = \nu_0={c\over 2\pi\rho}  
,\end{eqnarray}
where we evaluated $\sigma_{\rm s}$ at the  the fundamental harmonic
$\nu_{\rm B}=\nu_{\rm L}/\gamma$. 
We note that $\sigma_{\rm T} \propto 1/m^2_{\rm e}$, $\sigma_{\rm c}\propto 1/m_{\rm e}$
while $\sigma_{\rm S}$ is independent of the mass of the particle (at the fundamental frequency).

The absorption of curvature radiation can be particularly relevant for large values of the
magnetic fields, like those in the vicinity of the surface of neutron stars ad magnetars, where
strong synchrotron losses make the particles rapidly lose their initial pitch angle,
while leaving unaltered the momentum along the magnetic field lines.
Furthermore, along these lines, the particles could accelerate through
the process of magneto-centrifugally driven acceleration (see e.g.  Rieger \&  Mannheim  2000;  
Osmanov,  Rogava  \&  Bodo  2007;  Rieger  \& Aharonian 2008), using the spin of the neutron star.
In the self-absorbed regime, this acceleration process is not limited by 
radiative cooling (the absorption balances the emission), and this makes the particle
distribution to have a relatively large low energy cut-off $\gamma_{\rm min}$. 

The other crucial difference between the synchrotron and curvature cross sections is 
the impossibility of maser for the curvature radiation. 
This result is particularly relevant for the origin of the huge brightness temperature seen
in pulsars 
(see e.g. the review by Usov 2000)
and FRBs ($T_{\rm B}>10^{34}$ K; see e.g. the review by Katz 2016) 
requiring coherent radiation.
If the millisecond pulses seen in FRB are indeed due to curvature radiation, then
coherence cannot be associated with maser action without taking into account 
effect such as Cherenkov--curvature or Cherenkov--drift instabilities 
(Lyutikov, Machabeli \& Blandford 1999, Lyutikov, Blandford \& Machabeli 1999). 
Whenever the conditions for their action are not satisfied,
coherence must necessarily be due to the bunching of particles contained in one wavelength
(e.g. Kumar, Lu \& Bhattacharya 2017) and not to a maser.
This, together with the large absorption cross section derived in this work, poses the
problem of how to avoid the self-absorption of the produced curvature radiation, 
even in the coherent case.
We plan to investigate possible solutions to this problem in a future paper.

\section*{Acknowledgements}
We thank Sergio Campana, Luca Zampieri, Giancarlo Ghirlanda and Fabrizio Tavecchio  for discussions.
We thank  the  anonymous  referees  for critical  comments that helped to improve the paper.
NL acknowledges financial contribution from the ERC Starting Grant "MAGCOW", no.714196.
% We acknowledge financial contribution from the agreement ASI--INAF I/037/12/0
% (NARO 15) and from the CaRiPLo Foundation and the
% regional Government of Lombardia for the project ID 2014-1980 ``Science and technology
% at the frontiers of $\gamma$--ray astronomy with imaging atmospheric Cherenkov Telescopes".

% {\tt http://tools.asdc.asi.it}

% the National Aeronautics and Space Administration.
% This work made use of data supplied by the UK Swift
% Science Data Centre at the University of Leicester.

\vskip 1cm
% \noindent
% {\bf APPENDIX}
\appendix 

\section{Emissivity and absorption coefficient}

% \subsubsection{Power law distribution of particles}

The emissivity $\bar j(\nu)$ of particles distributed as a power law for a source with tangled magnetic field is
\begin{equation}
\bar j_{\rm c}(\nu) \, =\, {1\over 4\pi} \int p_c(\nu,\gamma) N(\gamma) d\gamma
,\end{equation}
where $N(\gamma) = N_0 \gamma^{-n}$ and $p_c(\nu,\gamma)$ is given by Eq. \ref{psingle}. 
The $4 \pi$ factor implies that we are assuming an isotropic emission. 
Using Eq. 35 in Westfold (1959)
\begin{eqnarray}
\int_0^\infty t^{\mu-1} {\int_t^\infty K_{a+1}(y)dy }{}dt  =  \frac{a+\mu}{\mu} \int_0^\infty t^{\mu-1} K_a(t)dt \\
\int_0^\infty t^{\mu-1} K_a(t)dt   =  2^{\mu-2}
\, \Gamma\left({ \mu+a\over 2 }\right) \,
\Gamma\left({ \mu-a\over 2}\right)
\label{int_xK}
\end{eqnarray} 
we have
\begin{eqnarray}
\bar j_{\rm c}(\nu) &=&  D(n) \cdot {e^2 N_0\over  \rho} \left( {\nu\over \nu_0} \right)^{-(n-2)/3}  
\nonumber \\
D(n) &\equiv &{ 3^{(n-2)/3} \over 8 \pi \sqrt{3} } {n+3\over n+1} 
\Gamma\left({n-1\over 6}\right) \Gamma\left({n+3\over 6}\right)
\label{emiss}
,\end{eqnarray}
which coincides with Cocke \& Pacholczyk (1975) apart from the factor $(n+3)/(n+1)$,
that is  missing from their Eq. 5.

Comparing with synchrotron emissivity made by the same particle distribution,
$\bar j_{\rm s}(\nu) \propto \nu^{-(n-1)/2}$, we can see that curvature emissivity 
has a steeper spectrum.

% ============================

The absorption coefficient for curvature radiation can be simply derived by
integrating the cross section over the energy distribution of electrons.
For a power law,  $N(\gamma)= N_0\gamma^{-n}$ between
$\gamma_{\rm min}\sim 1$ and $\gamma_{\rm max}\gg 1$, we have
\begin{equation}
\alpha_{\rm c}(\nu) \, =\, \int \sigma_{\rm c}(\nu,\gamma) N(\gamma) d\gamma  =
C(n){e^2 \over mc^2}\rho  N_0 \left( {\nu\over \nu_0} \right)^{-(n+5)/3} 
%\nonumber \\
%C(n) &= &{3^{(2n+1)/6}\over 48} \cdot n \,{n+4 \over n+2} \,
%\Gamma \left({n\over 6}\right) \, \Gamma \left({n+4\over 6}\right) 
\label{alpha}
,\end{equation}
where $C(n)$ is given in Eq. \ref{cn}, and where we 
used again Eq.~\ref{int_xK} (see also Eq. 11.4.22 of Abramowitz \& Stegun, 1975).
%
% \textbf{$C(n)$ is positive for all values of $n$. 
% Typical values for $C(n)$ are $C(1)=3.4$, $C(2)=6.7$, $C(3)=13.0$ and $C(4)=25.1$.}  

The source function is
\begin{eqnarray}
S_{\rm c}(\nu) &\equiv& {\bar j_{\rm c}(\nu)\over \alpha_{\rm c}(\nu)} 
= B(n) {mc^2\over \rho^2} \left( \nu \over \nu_0 \right)^{7/3}  \nonumber \\
B(n) &=& {D(n) \over C(n)} 
 = {2\over 3^{1/3}\pi n} {(n+3)(n+2) \over (n+1)(n+4)}\, {\Gamma [{n-1\over 6}] \, \Gamma [{n+3\over 6}] \over \Gamma [{n\over 6}] \, \Gamma [{n+4\over 6}]}    
\label{source}
.\end{eqnarray}
We note the $\nu^{7/3}$ slope, unlike the self--absorbed synchrotron spectrum, 
that has a $\nu^{5/2}$ slope.

%--------------------------------------------------
\begin{figure} 
\vskip -2.9 cm
\hskip -0.5 cm
\includegraphics[width=9.8cm]{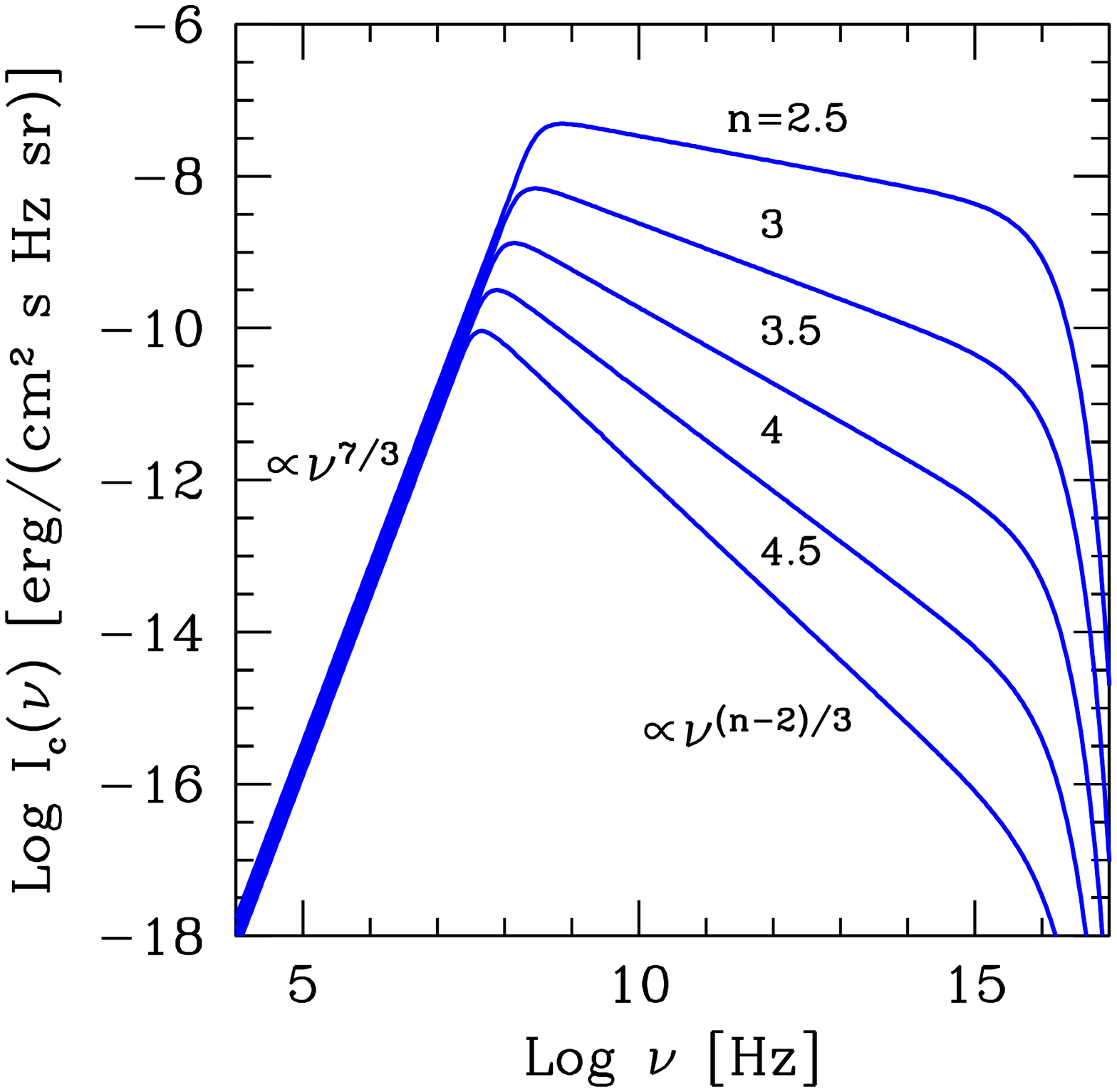} % ,width=12cm}
\vskip -2.8 cm
\caption{Specific intensity of curvature radiation for different spectral indexes $n$ of the electron distribution.
The other parameters are
$\rho=10^6$ cm, $R=10^4$ cm, $N_0=10^{15}$ cm$^{-3}$,
$\gamma_{\rm max}=2\times 10^4$.
Note the $\nu^{7/3}$ behaviour in the self--absorbed part.
} 
\label{spectrum}
\end{figure}
%--------------------------------------------------

%======================

Fig. \ref{spectrum}  shows how the specific intensity changes by changing the low energy cut-off
$\gamma_{\rm min}$ of the electron distribution.

%For $\gamma_{\rm min}>1$, the low frequency part of the self--absorbed
%spectrum scales as $\nu^2$ up to $ \sim\nu_0\gamma_{\rm min}^3$.
%When $\nu_0\gamma^3_{\rm min}>\nu_{\rm t}$, where $\nu_{\rm t}$ is the self-absorption frequency, also the thin %spectrum is produced by particles of energy $\gamma_{\rm min}$: in this case $I_{\rm c}(\nu) \propto \nu^{1/3}$.

\end{document}